\pdfoutput=1

\documentclass[11pt]{article}

\usepackage{ACL2023}

\usepackage{times}
\usepackage{latexsym}
\usepackage{booktabs}
\usepackage{graphicx}
\usepackage{xcolor, soul}
\sethlcolor{yellow}

\usepackage{multirow}

\usepackage{amsmath}

\graphicspath{ {./images/} }

\usepackage[T1]{fontenc}

\usepackage[utf8]{inputenc}

\usepackage{microtype}

\usepackage{inconsolata}

\title{The SVASR System for Text-dependent Speaker Verification (TdSV) AAIC Challenge 2024}

\author{
    Mohammadreza Molavi* \\
    Department of Computer Engineering \\
    Amirkabir University of Technology \\
    \texttt{mmdreza.molavi@aut.ac.ir} \\
    \And
    Reza Khodadadi* \\
    Department of Electrical Engineering\\
    Sharif University of Technology \\
    \texttt{khodadadi.reza@ee.sharif.edu}
}

\begin{document}

\renewcommand{\arraystretch}{1.5} 

\maketitle

\begin{abstract}

This paper introduces an efficient and accurate pipeline for text-dependent speaker verification (TDSV), designed to address the need for high-performance biometric systems. The proposed system incorporates a Fast-Conformer-based ASR module to validate speech content, filtering out Target-Wrong (TW) and Impostor-Wrong (IW) trials. For speaker verification, we propose a feature fusion approach that combines speaker embeddings extracted from wav2vec-BERT and ReDimNet models to create a unified speaker representation. This system achieves competitive results on the TDSV 2024 Challenge test set, with a normalized min-DCF of 0.0452 (rank 2), highlighting its effectiveness in balancing accuracy and robustness.

\textbf{Index Terms:} speaker verification, text-dependent, ASR,
speaker embedding fusion, TDSV 2024 Challenge

\end{abstract}

\section{Introduction}

Text-dependent speaker verification (TdSV) is a method of speaker authentication in which the speaker is required to utter a specific phrase or password, contrasting with text-independent verification (TiSV) that analyzes spontaneous speech or conversational audio. In text-dependent verification, the system leverages both the acoustic and linguistic properties of a predefined phrase to confirm the speaker's identity. 

This paper presents our work in a recent challenge on text-dependent speaker verification, organized by \cite{tdsvc2024plan}, provided a large dataset of native Persian speakers. The dataset consisted of a fixed set of phrases, including five sentences in Persian and five in English.

I-vector methods have seen extensive use in speaker verification over recent years. One notable technique employs bottleneck (BN) features extracted from triphone-state classifiers as input for i-vector extraction\cite{liu2015deep}. Unlike traditional approaches, BN-based i-vector extractors utilize Baum-Welch statistics derived from BN features, which effectively capture phonetic details and enhance the representation of textual content in speech. While this method is not entirely new, it remains highly competitive. For instance, \cite{lozano2020but} has shown that BN i-vectors can surpass x-vectors in performance when ample in-domain training data is available. However, a key drawback lies in the necessity of training a phonetic-aware deep neural network (DNN) on large-scale speech datasets to produce BN features, which adds significant complexity and cost to the deployment of speaker verification systems.

Multitask learning has become a common approach in TdSV, as it allows models to leverage both speaker and phoneme information for improved performance. In \cite{liu2021phoneme} a multitask learning framework was proposed to produce speaker embeddings enriched with phoneme-level knowledge. The architecture includes a shared frame-level encoder, a frame-level phoneme classifier, a speaker classifier, and a segment-level phoneme classifier. Speaker-phoneme multitask learning has demonstrated significant performance improvements on the RSR2015 dataset compared to traditional TDSV systems \cite{liu2021phoneme}. However, the approach relies on an automatic speech recognition (ASR) model to generate phoneme labels, which adds complexity and increases deployment costs.

Multitask learning can improve TDSV performance during fine-tuning. In \cite{han2022sjtushortdurationspeakerverification}, two strategies were explored: speaker $+$ phrase, which uses separate classification heads for speaker and phrase labels, and speaker $\times$ phrase, which treats utterances with the same speaker but different phrases as distinct classes using a single head. The speaker $+$ phrase approach outperformed speaker $\times$ phrase on the text-dependent task in SdSVC 2021, demonstrating the effectiveness of multitask fine-tuning.

During the competition, our initial strategy involved fine-tuning large audio models using the speaker $+$ phrase approach. Subsequently, we explored analyzing phrases through an ASR model while handling speaker voices independently. This approach delivered the best performance, forming the foundation of our proposed model. Details of the models are provided in the system overview section, with results discussed in the experiments section.

\section{Background}

\subsection{Task Definitions}

Text-dependent speaker verification is a key approach to identity verification, requiring that both the speaker's identity and a specific spoken phrase match the target criteria. Given a test speech file containing a designated phrase and enrollment data from the target speaker, the system determines if the target speaker indeed uttered the specified phrase. Unlike text-independent verification, where the focus is solely on speaker identity, text-dependent verification requires verifying the content of the speech as well. This creates a dual verification task, necessitating the confirmation of both the speaker and the spoken phrase.

In this task, a speaker encoder model must be trained on a large dataset that includes ten predefined phrases. For each speaker, an enrollment model is created using three repetitions of a specific phrase selected from this set. Verification is then performed by having the speaker repeat the same phrase once; based on this input, the system confirms or rejects the speaker's identity.

\subsection{Dataset and Metrics}

The main dataset for this challenge is the DeepMine dataset, gathered through crowdsourcing\cite{deepmine2018odyssey}\cite{deepmine2019asru}. Enrollment and test phrases are selected from a fixed set of ten phrases, comprising five in Persian and five in English. However, for target-wrong trials, some test utterances are drawn from free-text phrases. The in-domain training set consists of utterances from 1,620 speakers, with some speakers contributing only Persian phrases.

The main performance metric for this challenge is the normalized minimum Detection Cost Function (DCF), as outlined in SRE08. This function combines the miss and false alarm error probabilities into a weighted sum:
\begin{multline*}
C_{\text{Det}} = C_{\text{Miss}} \times P_{\text{Miss | Target}} \times P_{\text{Target}} \\
+ C_{\text{FalseAlarm}} \times P_{\text{FalseAlarm | NonTarget}} \times (1 - P_{\text{Target}})
\end{multline*}
where the parameters are set as 
\( C_{\text{Miss}} = 10 \), \( C_{\text{FalseAlarm}} = 1 \), and \( P_{\text{Target}} = 0.01 \). 
To normalize the DCF, it is divided by 0.1, representing the best possible cost if no processing 
were applied to the input data. Alongside the normalized minimum DCF (\(\text{minDCF}_{\text{norm}}^{0.01}\)), 
the Equal Error Rate (EER) is also used as a secondary evaluation metric.

\section{System Overview}

This section outlines the model architectures used for the task. Initially, we employed the speaker $+$ phrase approach, a dual-head strategy with separate heads for speaker and phrase classification to jointly optimize both aspects. Subsequently, we adopted a method that processes text independently of speaker information using a single speech recognition model. This latter approach proved more effective in our experiments, achieving superior results.

\subsection{Dual-head Strategy}
In this section, we employ three approaches. We detail each approach below

\textbf{ResNet34:}
We adopted ResNet34 as the backbone for our embedding extractor due to its proven efficiency in modeling complex data structures for speaker verification. Following the architecture introduced in \cite{zeinali2019but}, input features pass through an initial convolution layer, four residual blocks, and a statistical pooling layer that aggregates frame-level features into a segment-level representation. The final 256-dimensional fully connected layer produces fixed-length speaker embeddings. These embeddings were then used to train both the phrase classification and speaker classification heads in the speaker $+$ phrase approach.

\textbf{XEUS:}
We employed XEUS \cite{chen2024towards}, a Cross-lingual Encoder for Universal Speech, as our embedding extractor due to its extensive training on over 1 million hours of multilingual speech data across 4057 languages. XEUS significantly expands the language coverage of self-supervised learning models and enhances robustness through a novel dereverberation objective tailored for diverse multilingual datasets. The XEUS pre-trained weights were frozen during training, and its extracted embeddings were utilized to train both the phrase classification and speaker classification heads in the speaker $+$ phrase approach.

\textbf{Hybrid network with attention:}
Building on the assumption that XEUS contains rich information about both the speaker and the spoken content, we integrated a D-vector model \cite{wan2018generalized} as a speaker encoder and a pre-trained ResNet for phrase classification. Using these as queries in a cross-attention mechanism, we extracted embedding vectors, which were then concatenated to form a unified representation. This concatenated vector served as the final embedding, which was used to train both the phrase classification and speaker classification models.

\subsection{SVASR}
In this section, we have a independenet approach for modeling speaker and speech content. So we have a speaker verification with ASR (SVASR). 
\subsubsection{speaker embedding extractor models}
We used two speaker embedding extractor for train a speaker verification model in a text independent manner.

\textbf{W2V-BERT:}
After success of SSL models that trained with masked prediction loss like WavLM \cite{chen2022wavlm} and Hubert \cite{hsu2021hubert} in speaker verifiacation tasks, we used w2v-BERT model \cite{chung2021w2v}.
The w2v-BERT model, inspired by masked language modeling (MLM) in natural language processing, applies MLM for self-supervised speech representation learning. It combines MLM with contrastive learning, where contrastive learning transforms continuous speech into discrete tokens, and MLM helps the model learn contextualized speech representations by predicting masked tokens. Unlike other MLM-based speech models like HuBERT and vq-wav2vec \cite{baevski2020vqwav2vecselfsupervisedlearningdiscrete}, which use multi-stage or separate training processes, w2v-BERT achieves end-to-end optimization by jointly solving both tasks.

\textbf{ReDimNet:}
Reshape dimensions network (ReDimNet)  is an advanced neural network designed for extracting utterance-level speaker representations by reshaping dimensions between 2D feature maps and 1D signal representations, integrating both 1D and 2D blocks within a single framework \cite{yakovlev2024reshape}. Its unique topology preserves the channel-timestep-frequency volume, enabling efficient aggregation of feature maps. ReDimNet is scalable, with model variants ranging from 1 to 15 million parameters and computational costs between 0.5 and 20 GMACs, delivering state-of-the-art performance in speaker recognition while minimizing computational requirements and model size. To further improve accuracy, ReDimNet is trained on a challenging dataset, enhancing its performance in speaker recognition tasks.

For speaker verification, we propose a feature fusion approach that combines speaker embeddings extracted from wav2vec-BERT and ReDimNet models to form a unified speaker representation. The fusion method employs concatenation, where the embedding vectors from wav2vec-BERT and ReDimNet are first normalized and then combined with equal weighting.

\subsubsection{ASR}

In the text-dependent speaker verification task, we employ an Automatic Speech Recognition (ASR) system to filter out trials where the enrollment and test utterances originate from different phrases. For this, we use a FastConformer-based ASR model \cite{rekesh2023fastconformerlinearlyscalable}, implemented within the NeMo framework \cite{kuchaiev2019nemotoolkitbuildingai}. Fastconformer is a version of Conformer \cite{gulati2020conformer} model with reduced size of convolutional kernels and 2x subsampling rate.

Our starting checkpoint \footnote{\url{https://catalog.ngc.nvidia.com/orgs/nvidia/teams/nemo/models/stt_fa_fastconformer_hybrid_large}} has been  finetuned with Persian Mozilla Common Voice from english ASR model \footnote{\url{https://catalog.ngc.nvidia.com/orgs/nvidia/teams/nemo/models/stt_en_fastconformer_hybrid_large_pc}}.

\subsubsection{SVASR Pipline}
In this section, we describe the interaction between the ASR model and the Speaker Verification model, as outlined in the previous subsections. The ASR system is evaluated by recognizing phrases and classifying utterances based on their Character Error Rate (CER) relative to the reference phrases. The transcriptions generated by the ASR model are used to filter out Impostor-Wrong (IW) and Target-Wrong (TW) trials in the test dataset through punitive scoring, assigning them low scores. For speaker verification, cosine similarity is calculated between the speaker model and the extracted speaker vector, which serves as the final metric for speaker recognition. The entire process is illustrated in Figure 1.

\begin{figure}
	\centering
	\includegraphics[width=8cm, height=5cm]{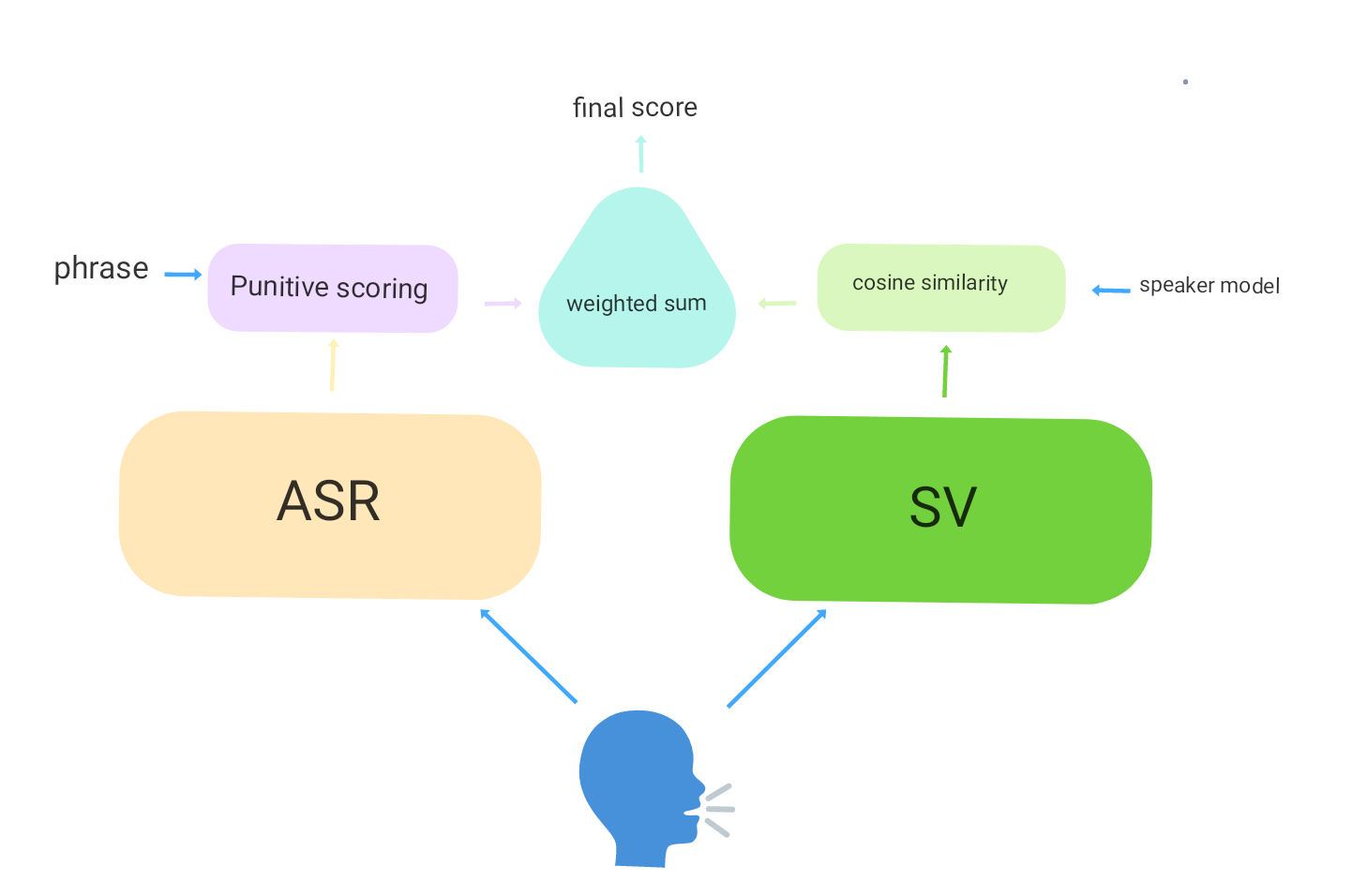}
	\caption{ Overview of the pipeline interaction between the ASR and Speaker Verification models, including the process of filtering IW and TW trials and scoring speaker verification using cosine similarity.}
    \vspace{-3mm}
	\label{fig:my_label}
\end{figure}

\section{Experimental Setup and Results}

In this section, we explain our training configuration and setup for each strategy and report our trained models`s  results on develop and test sets.

\subsection{Dual-head Strategy}

\subsubsection{Experimental Setup}

In the system overview section, we describe the Dual-head strategy models. All models were trained on the challenge dataset's training set using the AdamW optimizer with a learning rate of \(1 \times 10^{-3}\) and a batch size of 32. We used cross-entropy loss to train the ResNet model for phrase classification within the hybrid network, while AAM-softmax loss \cite{Wang_2018} was employed to train the ResNet34 and XEUS-based models in the dual-head strategy.

\subsubsection{Results}
We evaluated the performance of all models on the development set. Based on the results from the development set, we calculated the Log-Likelihood Ratios (LLR) for the test set trials using the best-performing model. Details are provided in Table~\ref{tab:dual_head_results}.

\begin{table}
    \centering
    \caption{Performance evaluation of all models on the development set, and results for the best model based on development set performance on the test set in terms of min-DCF}
    \setlength\tabcolsep{6pt}
     \begin{tabular}{ l | cc } 
    \toprule
    \toprule
    Model & Dev & Test  \\
    \midrule
     ResNet34 & 0.58 & -  \\ 
     XEUS & 0.43 & - \\
     Hybrid network with attention & 0.23 & 0.40 \\
     \bottomrule
    \bottomrule
     \end{tabular}
    \vspace{-3mm}
    \label{tab:dual_head_results}
\end{table}

\subsection{SVASR}

\subsubsection{Experimental Setup}

As outlined in the previous section, we trained the w2v-BERT and ReDimNet models separately. For w2v-BERT, we fine-tuned the TDNN layers, while the entire ReDimNet network was fine-tuned, both using the training set from the challenge dataset. Each model was optimized over 5 epochs using the AdamW optimizer with a learning rate of \(1 \times 10^{-5}\), incorporating square scheduling with 8000 steps and a batch size of 32. The training process employed AAM-softmax loss and SphereFace2 \cite{wen2022sphereface} loss.

The English ASR model \footnote{\url{https://catalog.ngc.nvidia.com/orgs/nvidia/teams/nemo/models/stt_en_fastconformer_hybrid_large_pc}} is initially pre-trained on the Persian Mozilla Common Voice dataset \footnote{\url{https://commonvoice.mozilla.org/fa/datasets}} and is then fine-tuned on an in-domain dataset consisting of five English and five Persian sentences. First, we had two ASR model. One for english and one for persian. Both models have same architecture. During fine-tuning separate models, only the decoder of the ASR model is updated, while the encoder remains frozen. To ensure uniform language processing and compatibility with a single tokenizer, the English sentences are transliterated into Persian script. Then, we fine-tuned entire ASR model with this new data.

To address potential bias due to the limited set of ten sentences, we enhance the ASR model by incorporating free-text utterances into the training data. These free-text utterances are transcribed using the pre-trained ASR model, and the resulting transcriptions are combined with the original dataset. The entire ASR model is then fine-tuned on this augmented dataset, using a self-training approach. In this stage we train our model for 3 epoch with learning rate \(5 \times 10^{-5}\) and AdamW optimizer. Also, improve generaliztion and robustness of model, SpecAugment \cite{park2019specaugment} technique is used for data augmentation. Training process employed Hybrid-CTC-Transducer loss for better and faster convergence.

\subsubsection{Results} 

As we explined before, ASR system is a front-enf of our pipeline. For realize real performance of our ASR, we compare result of our system on TC-vs-TW mode. Means that, we just select these trials and compute min-DCF on them.
We have evaluated our trained model on develop and test sets. For each experiment, a threshold is set for accepting a trial is target or imposter based on CER. In Table~\ref{tab:asr_results} we compare result of each ASR model.

After training the ASR model, it was integrated into the pipeline, and a series of experiments were conducted on the speaker models that were trained. The results revealed that models trained with SphereFace2 loss outperformed others. This is attributed to the loss function's ability to map speaker embeddings onto a spherical space, thereby maximizing the margin between different speakers. Subsequently, feature fusion via concatenation was employed to further enhance performance. For scoring, cosine similarity was calculated between the fused speaker model and the fused test utterance speaker embedding. Details are provided in Table~\ref{tab:svasr_results}.

\begin{table}[ht]
    \centering
    \caption{Performance comparison of the ASR proposed system on the challenge Dev-set and Test-set languages using normalized min-DCF as the evaluation metric.}
    \resizebox{81mm}{!}{ 
    \begin{tabular}{l | c c c} 
    \toprule
    \toprule
    Model & CER Threshold & Dev & Test \\
    \midrule
    Pre-trained & 0.4 & 0.0221 & 0.0545 \\ 
    Fine-tuned separately & 0.1 & 0.0089 & 0.0196 \\
    Fine-tuned (without free-text) & 0.02 & 0.0008 & 0.0026 \\
    Fine-tuned (with free-text) & 0.3 & \textbf{0.0001} & \textbf{0.0006} \\
    \bottomrule
    \bottomrule
    \end{tabular}
    }
    \vspace{-3mm}
    \label{tab:asr_results}
\end{table}

\begin{table}[h!]
  \caption{Performance comparison of SVASR system. Last three columns have format : Speaker model + ASR. PT means pre-trained SF2 means SphereFace2 and FT means finetuned. ft means free text usage}
  \footnotesize
  \label{t:LibriSpeech}
  \centering
  \resizebox{\columnwidth}{!}{%
  \begin{tabular}{ccrrr}
    \toprule
    \toprule
    \multirow{1}{*}{\bfseries Network} 
    & \multicolumn{1}{c}{\bfseries EER(\%)}  & \multicolumn{2}{c}{\bfseries min-DCF} \\
    \cmidrule(r){2-2} \cmidrule(r){3-4}
     & {\bfseries Test} & {\bfseries Dev} & {\bfseries Test} \\
    \midrule
    {ReDimNet(PT) + FT(Sep)} & 3.09 & 0.0550 & 0.0882 \\
    \midrule
    {W2V-BERT(AAM) + FT(Sep)}  & - & 0.0841 & -  \\
    \midrule
    \multirow{1}{*}{W2V-BERT(SF2) + FT(Sep)} 
    & - & 0.0526 & -  \\
    \midrule
    \multirow{1}{*}{ReDimNet(SF2) + FT(Sep)} & - & 0.0467 & -  \\
    \midrule
    \multirow{1}{*}{W2V-BERT(SF2) + FT(w.o ft)} & - & 0.0438 & -   \\
    \midrule
    \multirow{1}{*}{ReDimNet(SF2) + FT(w.o ft)} & 1.8 & 0.0351 & 0.0747  \\
    \midrule
    \multirow{1}{*}{Fusion Result + FT(w.o ft)} & 1.32 & 0.0242 & 0.0453   \\
    \midrule
    \multirow{1}{*}{Fusion Result + FT(with ft)} & \textbf{1.35} & \textbf{0.0232} & \textbf{0.0452}   \\
    \bottomrule
    \bottomrule
  \end{tabular}
  }
  \label{tab:svasr_results}
\end{table}

\section{Conclusion}

Based on the obtained results, we propose a solution for the challenge of text-dependent speaker verification where the input sentence is predetermined. Our approach involves employing an initial filtering stage based on an Automatic Speech Recognition (ASR) model. In this stage, the speech content is first validated, followed by the speaker verification process. 

In the second stage, speaker verification is refined by combining embeddings from two models. Speaker embeddings extracted using wav2vec-BERT are concatenated with those obtained from ReDimNet to form a unified speaker representation. This concatenated vector serves as the final speaker embedding, effectively leveraging the strengths of both models to enhance precision, particularly for challenging trials. 

For future work, we suggest integrating this system with foundational models while leveraging knowledge distillation techniques to create smaller, task-specific versions of these foundational models for each component.

\bibliography{custom}
\bibliographystyle{acl_natbib}

\end{document}